\documentstyle[12pt]{article}

\newcommand{\sect}[1]{\setcounter{equation}{0}\section{#1}}

\def\rf#1{(\ref{eq:#1})}
\def\lab#1{\label{eq:#1}}
\def\nonu{\nonumber}
\def\br{\begin{eqnarray}}
\def\er{\end{eqnarray}}
\def\be{\begin{equation}}
\def\ee{\end{equation}}

\def\lb{\lbrack}
\def\rb{\rbrack}

\def\({\left(}
\def\){\right)}

\newcommand{\ct}[1]{\cite{#1}}
\newcommand{\bi}[1]{\bibitem{#1}}
\def\lskip{\vskip\baselineskip\vskip-\parskip\noindent}
\relax

\def\v{\vert}

\newcommand{\sbr}[2]{\left\lbrack\,{#1}\, ,\,{#2}\,\right\rbrack}

\def\a{\alpha}

\def\bfn{{\bf n}}
\def\d{\delta}

\def\eps{\epsilon}

\def\h{{1\over 2}}
\def\l{\lambda}

\def\o{\over}

\def\pa{\partial}
\def\pr{\prime}

\def\ra{\rightarrow}
\def\s{\sigma}

\def\th{\theta}

\def\tp0{\Theta_{+}^{(0)}}
\def\tm0{\Theta_{-}^{(0)}}

\def\u2{\mid u\mid^2}

\def\cf{{\cal F}}
\def\cg{{\cal G}}

\def\ck{{\cal K}}
\def\cl{{\cal L}}

\font \msb=msbm10 scaled \magstep1
\newcommand{\IR}{\mbox{\msb R} }

\def\one{\hbox{{1}\kern-.25em\hbox{l}}}
\def\0#1{\relax\ifmmode\mathaccent''7017{#1}%
B        \else\accent23#1\relax\fi}

\def\PRL#1#2#3{{\sl Phys. Rev. Lett.} {\bf#1} (#2) #3}
\def\NPB#1#2#3{{\sl Nucl. Phys.} {\bf B#1} (#2) #3}

\def\PLB#1#2#3{{\sl Phys. Lett.} {\bf #1B} (#2) #3}

\def\LMP#1#2#3{{\sl Letters in Math. Phys.} {\bf #1} (#2) #3}

\begin{document}

\begin{titlepage}
\vspace*{-2 cm}
\noindent
February, 1999 \hfill{IFT-P.019/99}\\
hep-th/9902141 \hfill{UICHEP-TH/99-1}

\vskip 3cm

\begin{center}
{\Large\bf  Toroidal solitons in ${\bf 3+1}$  dimensional\\ 
  integrable theories}
\vglue 1  true cm
H. Aratyn$^1$ , L.A. Ferreira$^2$ and A.H. Zimerman$^2$\\

\vspace{1 cm}
$^1${\footnotesize Department of Physics \\
University of Illinois at Chicago\\
845 W. Taylor St., Chicago, IL 60607-7059}

\vspace{1 cm}

$^2${\footnotesize Instituto de F\'\i sica Te\'orica - IFT/UNESP\\
Rua Pamplona 145\\
01405-900, S\~ao Paulo - SP, Brazil}\\

\medskip
\end{center}

\normalsize
\vskip 0.2cm

\begin{abstract}

We analyze the integrability properties of models defined on the symmetric
space $SU(2)/U(1)$ in $3+1$ dimensions, using a recently proposed approach for
integrable theories in any dimension.  
We point out the key ingredients for a theory to possess an infinite number
of local conservation laws, and discuss classes of models with such
property.  
We propose a $3+1$-dimensional, relativistic  invariant 
field theory possessing a toroidal soliton solution carrying a unit of
topological charge given by the Hopf map.  
Construction of the action is
guided by the requirement that the  energy of static configuration should be
scale invariant. The solution is constructed exactly. The model possesses an
infinite number of local conserved currents. 
The method is also applied to the Skyrme-Faddeev model, and integrable 
submodels are proposed. 

\end{abstract}
\end{titlepage}

\sect{Introduction} 

In this paper we consider scalar field theories in $3+1$ dimensions, defined 
on $S^2$, or equivalently on the symmetric space $SU(2)/U(1)$. One of the
motivations to study such theories is that some of them present topological
solitons. The requirement of finite energy for static
configurations imposes, in general,  that the fields should  be constant at
spatial infinity. Consequently, for such purpose space can be taken to be
$S^3$, and the solutions define a mapping $S^3 \ra S^2$. The topological
charges carried by the solitons are then determined 
by a Hopf map. That differs from the case of magnetic monopoles, for
instance, where the charges are winding numbers of the map $S^2\ra S^2$. The
Hopf index is given by the linking number of  the pre images of a given pair
of points of $S^2$. As a consequence, the solitons tend to have string like
configurations, and those with charge unity to have a toroidal shape. 

Conventional wisdom of two dimensional soliton physics holds that 
the existence of solitons  is linked to the notion of integrability. 
The reasoning is that the high degree of symmetries underlying the infinite set
of conserved quantities accounts for the conspiracy among the degrees of
freedom, necessary for the appearance of solitons. 
We give indications here that also in the setup of higher dimensional
integrable models  the solitons appear in theories with infinite number of
conserved quantities. 

We analyze the integrability properties of those scalar theories using the
approach of \ct{afg}, which generalizes the concept of zero 
curvature in two dimensions to theories defined in a space-time of any
dimension. Those ideas are reviewed in section \ref{sec:integ}. In section
\ref{sect:su2u1} we define the models we are interested 
in by presenting their zero curvature representation. Then we discuss the
conditions the model has to satisfy to contain an infinite number of
local conserved currents. In section \ref{sec:examples} we study some
examples of such integrable theories. One of the important ingredients to have
soliton solutions, is that the energy should be stable under scaling of the
space variables (Derrick's theorem). In section \ref{sect:nicemodel} we
introduce a model where the energy for static configurations is invariant
under such scalings. We then  construct the exact solution for a soliton
carrying one unity of topological charge. In section \ref{sect:sf} we discuss
the integrability of the Skyrme-Faddeev model \ct{faddeev}, and propose a
submodel of it which possesses an infinite number of conserved currents.

\sect{Integrability in any dimension}
\label{sec:integ}

As we said, we shall analyze the integrability properties of the models
considered in this paper  using the approach of \ct{afg}. The main idea there  
is to generalize the zero curvature condition in two dimensions guided by the 
fact that it embodies conservation laws. Indeed, the flatness condition for
the Lax operators $A_{\mu}$ implies that its path ordered integral is path
independent, as long as the end point are kept fixed. For a closed path that
leads to a Gauss type  law and so, conserved quantities. Therefore, the
central idea in \ct{afg} to bring such concepts to higher dimensions, is 
to introduce quantities integrated over hypersurfaces and to find the 
conditions for them to be independent of deformations of the hypersurfaces 
which keep their boundaries fixed. Such an approach  certainly leads to 
conservation laws in a manner very similar to the two dimensional case. 
However, the main problem of that is how to introduce non-linear zero 
curvatures keeping things as local as possible. The way out is to introduce
connections to allow for parallel transport. 

The zero curvature obtained in \ct{afg} is in general non local but 
there are interesting conditions under which it becomes local. The structures
underlying those conditions  involve a Lie algebra $\cg$  and a representation 
$R$ of it. Then one introduces the nonsemisimple Lie algebra $\cg_R$ as
\br
\lb T_a \, , \, T_b \rb &=& f_{ab}^c T_c \nonu\\
\lb T_a \, , \, P_i \rb &=& P_j R_{ji}\( T_a\) \nonu\\
\lb P_i \, , \, P_j \rb &=& 0
\lab{rt}
\er
where $T_a$ constitute a basis of $\cg$ and $P_i$ a basis for the abelian
ideal $P$ (representation space). The fact that $R$ is a matrix
representation, i.e.
\begin{equation}
\lb R\( T_a \) \, , \, R\( T_b\)  \rb =  R\( \lb T_a \, , \, T_b \rb \)
\lab{rep}
\end{equation}
follows from Jacobi identities.

In $(3+1)$ dimensions, which is the case of interest here, one then introduces
a  connection $A_{\mu}$  belonging to $\cg$ and  a rank $3$  
antisymmetric tensor $B_{\mu\nu\rho}$ belonging to $P$, i.e.
\begin{equation}
A_{\mu} = A_{\mu}^a T_a \; , \qquad B_{\mu\nu\rho} = 
B_{\mu\nu\rho}^i P_i
\end{equation}

Then the {\em local} zero curvature conditions are given by 
\be
D_{\l} B_{\mu\nu\rho} - D_{\mu} B_{\nu\rho\l}  + D_{\nu} B_{\rho\l\mu} -  
D_{\rho} B_{\l\mu\nu} = 0 
\lab{lzc1a}
\ee
and
\be
F_{\mu\nu} \equiv  [\partial_{\mu} + A_{\mu} , \partial_{\nu} + A_{\nu} ]=0 
\lab{lzc2}
\ee
where we have introduced  the covariant derivative
\be
D_{\mu} \cdot \equiv \partial_{\mu}\, \cdot  + [A_{\mu} \, , \, \cdot \, ]
\lab{covder}
\ee

Introducing the dual of $B_{\mu\nu\rho}$ as
\be
{\tilde B}^{\mu} \equiv {1\o 3!} \, 
\varepsilon^{\mu \nu \rho \l} \, B_{\nu\rho\l}
\lab{dual}
\ee
one can write \rf{lzc1a} as
\be
D_{\mu} {\tilde B}^{\mu} = 0 
\lab{lzc1b}
\ee
The relations \rf{lzc2} and \rf{lzc1b} constitute the {\em local}
generalization to higher  
dimensions of the zero curvature condition in two dimensions. They lead to
local conservation laws. Indeed, since the connection $A_{\mu}$ is flat it can
be  written as 
\be
A_{\mu} = -\partial_{\mu} W \, W^{-1}
\lab{puregauge}
\ee
and consequently \rf{lzc1b} implies that the currents 
\be
J_{\mu} \equiv  W^{-1}\, {\tilde B}^{\mu} \, W 
\lab{currents}
\ee
are conserved:
\be
\partial_{\mu} \, J^{\mu} = 0
\lab{conserv}
\ee

The zero curvature conditions \rf{lzc1a} and \rf{lzc2}  are invariant under
the gauge transformations
\br
A_{\mu} &\ra & 
g \, A_{\mu} \, g^{-1} - \pa_{\mu} g \, g^{-1} \nonu\\
B_{\mu\nu\rho} &\ra &  
g \, B_{\mu\nu\rho} \, g^{-1} 
\lab{gauge}
\er
and 
\br
A_{\mu} &\ra & A_{\mu} \nonu\\
B_{\mu\nu\rho}  &\ra & B_{\mu\nu\rho} + 
D_{\mu} \a_{\nu\rho} + D_{\nu} \a_{\rho\mu} + D_{\rho} \a_{\mu\nu}
\lab{newgauge}
\er
In \rf{gauge} $g$ is an element of the group obtained by exponentiating the 
Lie algebra $\cg$. The transformations \rf{newgauge} are symmetries of 
\rf{lzc1a} and \rf{lzc2} as a consequence of the fact that the connection
$A_{\mu}$ is  
flat, i.e. $\lb D_{\mu} \, , \, D_{\nu}\rb = 0$. In addition, the parameters 
$\a_{\mu\nu}$ take values in the abelian ideal $P$. 

The currents \rf{currents} are invariant under the transformations 
\rf{gauge}, while under \rf{newgauge} they transform as 
\be
J_{\mu} \ra J_{\mu} + 
\varepsilon_{\mu\nu\rho\l} \partial^{\nu}\( W^{-1} \, 
\a^{\rho\l}\, W \) 
\lab{curtransf}
\ee

The transformations \rf{gauge} and \rf{newgauge} do not commute and their 
algebra is isomorphic to the non-semisimple algebra $\cg_R$ introduced 
in \rf{rt}.

\sect{Integrable models on ${\bf SU(2)/U(1)}$}
\label{sect:su2u1}

The models we shall be considering involve scalar fields living on the two
dimensional sphere $S^2$, and we will denote them as  $\bfn = \( n_1, n_2,
n_3\)$, with $\bfn^2=1$.  Alternatively, one can use the stereographic
projection of $S^2$ and work with two unconstrained scalar fields, which we
shall choose to constitute a complex scalar field $u$ related to $\bfn$ by
\be
{\bf n} = {1\o {1+\mid u\mid^2}} \, \( u+u^* , -i \( u-u^* \) , \u2 -1 \)
\lab{stereo}
\ee

The sphere $S^2$ can be mapped in a one-to-one manner into the symmetric space
$SU(2)/U(1)$ and we 
shall explore that fact to construct the local zero curvature conditions 
\rf{lzc1a} and \rf{lzc2}\footnote{The sphere $S^2$ is diffeomorphic to
$SO(3)/SO(2)$, and $SO(3) \sim SU(2)/Z_2$. The elements of the subgroup $U(1)$
in $SU(2)/U(1)$ are given by $g\(\theta\)=\exp \( i \theta T_3\)$. Therefore,
$g\(0\)= \one$ and $g\( 2\pi\) = -\one$, and such $U(1)$ has twice as many
elements as $SO(2)$. In fact, $SO(2) \sim U(1)/Z_2$, and so the points of $S^2$
and $SU(2)/U(1)$ are in one-to-one correspondence.}. The $U(1)$ is the
subgroup invariant under the 
involutive automorphism of $SU(2)$
\be
\s \( T_3\) = T_3 \qquad \qquad \s\( T_{\pm} \) = - T_{\pm}
\lab{autom}
\ee
where $T_3$, $T_{\pm}$ are the generators of $SU(2)$
satisfying\footnote{Actually, these are the generators of $SL(2)$. The
generators of $SU(2)$ are $T_i$, $i=1,2,3$ with $T_{\pm} = \h \( T_1 \pm i
T_2\)$.}   
\br
\lb T_3 \, , \, T_{\pm} \rb &=& \pm  \, T_{\pm} \; , \qquad
\lb T_{+} \, , \, T_{-} \rb =  2\, T_3
\lab{sl2alg}
\er
The automorphism \rf{autom} is inner and given by 
\be
\s \( T\) \equiv e^{i\pi T_3} \, T \, e^{-i\pi T_3}
\lab{inneraut}
\ee

The elements of $SU(2)/U(1)$ can be parametrized by the variable
$x\( g \) \equiv g \s \( g\)^{-1}$,  $g \in SU(2)$, 
since $x\( g \)=x\( g k\)$ with $k\in U(1)$. In addition one has that 
$\s \( x\) = x^{-1}$. We then choose the group element $W$ in \rf{puregauge}
to be of the form of the variable $x\( g \)$, i.e.
\be
W\equiv  e^{iuT_+}\, e^{\varphi T_3}\, e^{iu^*T_-} 
\lab{w}
\ee 
where $\varphi =  \ln (1 + \u2 )$. In the defining (spinor) representation of
$SU(2)$ one has
\br
R^{(1/2)}\( W\)= {1\over {\sqrt{1 + \mid u\mid^2}}}\, \( 
\begin{array}{cc}
1 & i u \\
i u^* & 1
\end{array} \)
\lab{gaugefixedg}
\er
Indeed, one can check that $\s \( W\) = W^{-1}$, using in \rf{inneraut} that
\br
R^{(1/2)}\( e^{i\pi T_3}\) =  
\( 
\begin{array}{cc}
i & 0 \\
0  & -i
\end{array} \)
\er

We then introduce the potentials 
\br
A_{\mu} &=& -\pa_{\mu} W \, W^{-1} \nonu\\
 &=& {1\o{ 1+\mid u \mid^2 }} \(  -i \pa_{\mu} u \, T_{+}
-i \pa_{\mu} u^* \, T_{-} +  \( u \pa_{\mu} u^* - u^* \pa_{\mu} u \) \, T_3 \)
\lab{intpota}\\
{\tilde B}_{\mu} &=& {1 \o{ 1+\mid u \mid^2}}
 \(  \ck_{\mu}  \, P_{1}^{(1)} -  \ck_{\mu}^* \, P_{-1}^{(1)} \)
\lab{intpotb}
\er
where $\ck_{\mu}$ is a functional of the fields $u$ and $u^*$ and
their derivatives. In addition, $P_{\pm 1}^{(1)}$ stand for the states of
eigenvalues $\pm 1$ of $T_3$ in the triplet representation of
$SU(2)$. According, to \rf{rt} they are generators of the abelian subalgebra
$P$ of $\cg_R$. Here we give the commutation relations for any spin-$j$
representation ($m=-j,-j+1, \ldots , j-1, j$) 
\br
\lb T_3\, , \, P_{m}^{(j)} \rb &=&  m\,  P_{m}^{(j)}
\lab{sl2pa}\\
\lb T_{\pm}\, , \, P_{m}^{(j)} \rb &=&
\sqrt{j(j+1) - m(m\pm 1)} \,\,\, P_{m\pm 1}^{(j)}
\lab{sl2pb}\\
\lb P_{m}^{(j)}\, , \, P_{m^{\pr}}^{(j^{\pr})} \rb &=& 0
\lab{sl2pc}
\er

Obviously, the zero curvature relation \rf{lzc2} is trivially satisfied 
because we have chosen $A_{\mu}$ of the pure gauge form. So, it does not
impose any condition on the fields $u$ and $u^*$. 

Requiring, that
\be
{\rm Im} \( \ck^{\mu} \pa_{\mu} u^* \) = 0
\lab{req1}
\ee
one obtains that the zero curvature condition \rf{lzc1b} implies that 
\be
\( 1 + \u2 \) \pa^{\mu} \ck_{\mu} - 2 u^* \ck_{\mu} \pa^{\mu} u = 0 
\lab{eqmotgen}
\ee
together with its complex conjugate equation. 

According to \rf{currents} and \rf{conserv} one gets three conserved currents
corresponding to three states of the triplet representation. They are given by
\be
J_{\mu}^{(1)} = \sum_{m=-1}^{1} \, J_{\mu}^{(1,m)} P_{m}^{(1)}
\ee
with 
\be
J_{\mu}^{(1,1)} = 
{{{\ck}_{\mu} + {\ck}^*_{\mu}\,{{{u}}^2}}\over 
   {{{\left( 1 + \u2 \right) }^2}}} \qquad 
J_{\mu}^{(1,0)}=
{{i\,{\sqrt{2}}\,\left( {\ck}^*_{\mu}\,{u} - 
       {\ck}_{\mu}\,{u}^* \right) }\over 
   {{{\left( 1 + \u2 \right) }^2}}} \qquad 
J_{\mu}^{(1,-1)}= - {J_{\mu}^{(1,1)}}^* 
\lab{3curr}
\ee

Notice that the condition \rf{req1} implies that the term in the direction of
$P_{0}^{(1)}$ vanishes, i.e.  
$\sbr{ \pa_{\mu} u \, T_{+} + \pa_{\mu} u^* \, T_{-}}{ \ck_{\mu}  \,
P_{1}^{(1)} -  \ck_{\mu}^* \, P_{-1}^{(1)}}=0$. Therefore, the equations of
motion \rf{eqmotgen} are determined only by the way that ${\tilde B}_{\mu}$
transforms under the $U(1)$ subgroup generated by $T_3$. In fact, 
${\tilde B}_{\mu}$ contains two irreducible representations of $U(1)$ which
are the singlets $P_{\pm 1}^{(1)}$ of charges $\pm 1$. That means that if we
change the  representation of $SU(2)$ where ${\tilde B}_{\mu}$ lives, we do
not change  the equations of motion if ${\tilde B}_{\mu}$ still 
transforms under the same two singlets of the $U(1)$ subgroup. What can
happen is that the commutator of the $T_{\pm}$ part of $A_{\mu}$ does not
commute with ${\tilde B}_{\mu}$ anymore, and then we get some additional
equations which should be considered as constraints on the
model. Consequently, we would be dealing with submodels of the original
theory. See ref. \ct{fl} for a detailed discussion on that.

One way of implementing these ideas is as follows. Any integer spin-$j$ 
representation of $SU(2)$  possesses a
charge zero singlet of the $U(1)$ generated by $T_3$, which is 
$P_{0}^{(j)}$. Therefore, if one considers representations of $SU(2)$ which
are tensor products of these representations, one obtains several singlets
of $U(1)$ transforming like $P_{\pm 1}^{(1)}$, which are given by tensor
products of  $P_{\pm 1}^{(j)}$ with copies of $P_{0}^{(j)}$. For instance one
has 
\be
\sbr{1\otimes T_3 + T_3\otimes 1}{P_{0}^{(j^{\pr})}\otimes P_{\pm 1}^{(j)}}= 
\pm \( P_{0}^{(j^{\pr})}\otimes P_{\pm 1}^{(j)}\) 
\lab{tensorrs}
\ee

Therefore, for the case of the tensor product of $n$ integer spin 
representations,  one introduces the potentials 
\br
A_{\mu}^{(j_1\ldots j_n)} &\equiv&  \sum_{l=0}^{n-1} 
\(\otimes 1\)^l \otimes  \( A_{\mu}^{\pm} + A_{\mu}^3\) \(\otimes 1\)^{n-l-1} 
\equiv A_{\mu}^{(j_1\ldots j_n,\pm )} + A_{\mu}^{(j_1\ldots j_n,3)}
\lab{cosetpottensor}\\
{\tilde B}_{\mu}^{(j_1\ldots j_n)} &\equiv&    \sum_{l=0}^{n-1}
  {\tilde P}_{0}^{(j_1)}\otimes {\tilde P}_{0}^{(j_2)} \otimes \ldots 
{\(  \ck_{\mu}  \, P_{1}^{(j_l)} - 
 \ck_{\mu}^* \, P_{-1}^{(j_l)} \)\o {1+\mid u \mid^2}} 
\otimes {\tilde P}_{0}^{(j_{l+1})} \ldots \otimes {\tilde P}_{0}^{(j_{n})} 
\nonu 
\er
with $j_l$, $l=1,2, \ldots n$, being positive integers numbers, and where we
have rescaled the zero charge singlets as
\be
{\tilde P}_{0}^{(j_{l})} \equiv  P_{0}^{(j_{l})}/ \sqrt{j_l(j_l+1)}
\lab{rescale}
\ee
In addition, we have denoted 
\br
A_{\mu}^{\pm}&\equiv& {1\o{ 1+\mid u \mid^2 }} \(  -i \pa_{\mu} u \, T_{+}
-i \pa_{\mu} u^* \, T_{-}\) \nonu\\
A_{\mu}^3 &\equiv& {1\o{ 1+\mid u \mid^2 }} 
 \(  u \pa_{\mu} u^* - u^* \pa_{\mu} u \) \, T_3 
\er

The zero curvature condition \rf{lzc2} for these potentials is still trivially
satisfied because $A_{\mu}^{(j_1\ldots j_n)}$ is of the pure gauge form. The
condition  
\rf{lzc1b} can be split in two terms: 
\be
\pa^{\mu} {\tilde B}_{\mu}^{(j_1\ldots j_n)} + 
\sbr{A_{\mu}^{(j_1\ldots j_n,3)}}{{\tilde B}_{\mu}^{(j_1\ldots j_n)}} = - 
\sbr{A_{\mu}^{(j_1\ldots j_n,\pm )}}{{\tilde B}_{\mu}^{(j_1\ldots j_n)}}
\lab{brokenzc}
\ee
The l.h.s. of such equation vanishes as a consequence of the equations of
motion \rf{eqmotgen}. 

We now have, using \rf{req1}, that 
\be
\sbr{A_{\mu}^{(j_1\ldots j_n,\pm )}}
{\ck^{\mu}\, P_{1}^{(j_l)} -  {\ck^{\mu}}^* \, P_{-1}^{(j_l)}} = 
-i \sqrt{j_l(j_l+1) - 2} \; \( \ck^{\mu} \pa_{\mu} u \, P_{2}^{(j_l)} - 
{\ck^{\mu}}^* \pa_{\mu} u^* \, P_{-2}^{(j_l)}\) 
\lab{apmk}
\ee
In addition 
\be 
\sbr{A_{\mu}^{(j_1\ldots j_n,\pm )}}{{\tilde P}_{0}^{(j_l)}} = -i  \; 
\( \pa_{\mu} u \,  P_{1}^{(j_l)} + \pa_{\mu} u^* \,  P_{-1}^{(j_l)}\)
\lab{apmpo}
\ee

Now, let us analyze the r.h.s. of \rf{brokenzc}. Consider the terms containing
commutators of 
$A_{\mu}^{(j_1\ldots j_n,\pm )}$ with ${\tilde P}_{0}^{(j_l)}$ and 
${\tilde P}_{0}^{(j_m)}$
($l<m)$). Then one gets, using \rf{apmpo}, the terms 
\br
& &{\tilde P}_{0}^{(j_1)} \otimes \ldots 
\( -i\) \( \pa_{\mu} u \,  P_{1}^{(j_l)} + \pa_{\mu} u^* \,  P_{-1}^{(j_l)}\) 
\otimes \ldots 
{\(  \ck_{\mu}  \, P_{1}^{(j_m)} - 
 \ck_{\mu}^* \, P_{-1}^{(j_m)} \)\o {1+\mid u \mid^2}} 
\otimes  \ldots \otimes {\tilde P}_{0}^{(j_{n})}
\nonu\\
&+& 
{\tilde P}_{0}^{(j_1)} \otimes \ldots 
{\(  \ck_{\mu}  \, P_{1}^{(j_l)} - 
 \ck_{\mu}^* \, P_{-1}^{(j_l)} \)\o {1+\mid u \mid^2}}  
\otimes \ldots 
\( -i\) \( \pa_{\mu} u \,  P_{1}^{(j_m)} + \pa_{\mu} u^* \,  P_{-1}^{(j_m)}\)
\otimes  \ldots \otimes {\tilde P}_{0}^{(j_{n})}
\nonu 
\er

Therefore, taking into account \rf{req1}, one observes that if one imposes the
constraint
\be
\ck_{\mu} \,\pa^{\mu} u = 0
\lab{constraint}
\ee
those two terms cancel. The same constraint cancels the terms involving the
commutator \rf{apmk}. Therefore, the r.h.s. of \rf{brokenzc} vanishes. 

Consequently, the zero curvature conditions \rf{lzc2} and \rf{lzc1b} for the
potentials \rf{cosetpottensor} lead to the equations of motion \rf{eqmotgen}
and the constraint \rf{constraint}. And so, they lead to the submodel defined
by equations 
\be
\pa^{\mu} \ck_{\mu} = 0 \qquad \qquad  \ck_{\mu} \,\pa^{\mu} u = 0
\lab{subgen}
\ee

According to \rf{currents} and \rf{conserv} one obtains the conserved currents 
\br
J_{\mu}^{(j_1\ldots j_n)} &=& \( W^{-1}\otimes \ldots \otimes W^{-1}\) 
{\tilde B}_{\mu}^{(j_1\ldots j_n)} 
\( W\otimes \ldots  \otimes W\) \nonu\\
&\equiv& \sum_{l=1}^n \sum_{m_l=-j_l}^{j_l} J_{\mu}^{(j_1\ldots j_n),
(m_1\ldots m_n)} \,  
P_{m_1}^{(j_1)}\otimes \ldots  \otimes P_{m_n}^{(j_n)}
\lab{infcur1}
\er
Therefore, one gets $\prod_{l=1}^{n} \( 2 j_l + 1\)$ currents. However, since
$n$ and $j_l$ can be any positive integer number, such  submodel contains an 
infinity of conserved currents. 
All such currents are linear in $\ck_{\mu}$ and $\ck_{\mu}^*$, with the
coefficients being functionals of $u$ and $u^*$. 

Notice that any current of the form 
\be
J_{\mu} \equiv  \ck_{\mu}\, {\d G \o \d u} - 
 \ck_{\mu}^*\, {\d G \o \d u^*}
\lab{infcur2}
\ee
with $G$
being any functional of $u$ and $u^*$ only (no derivatives), are conserved 
as a consequence of \rf{req1} and \rf{subgen}. We have checked that for the
case where all $j_l$'s 
are equal to $1$, the currents \rf{infcur1} are of the form \rf{infcur2}
\ct{fl}.

\sect{Examples}
\label{sec:examples}

The methods discussed above can be used to construct {\em integrable} models
with an infinite number of conserved currents in a space-time of any
dimension. The example of $CP^1$ in $(2+1)$ was discussed in \ct{afg}, and
corresponds to the choice $\ck_{\mu} \ra \pa_{\mu}u$. Examples involving
other symmetric spaces (or homogeneous spaces) were also considered in
\ct{fl,joaq,fujii}.

A particular class of models can be constructed using the quantity 
\be
K_{\mu} = \( \pa^{\nu} u^* \pa_{\nu} u \) \, \pa_{\mu} u - 
\(\pa_{\nu} u \)^2 \, \pa_{\mu} u^*
\lab{kmu}
\ee
since it automatically satisfies
\be
K_{\mu} \pa^{\mu} u = 0
\lab{reqk1}
\ee

In addition one has that
\be
{\rm Im}\( K_{\mu} \pa^{\mu} u^*\) = 0
\lab{reqk2}
\ee

Therefore, if $\cf$ is any real functional of $u$, $u^*$ and their
derivatives, it follows that the choice 
\be
\ck_{\mu} \ra \cf\( u\)  K_{\mu}
\ee
satisfies \rf{req1} and \rf{constraint}, and consequently leads to a class of
models defined by the equations of motion (see \rf{subgen})
\be
\pa^{\mu} \( \cf\( u\)  K_{\mu} \) = 0 
\ee
and possessing an infinite number of local conserved currents given by 
\rf{infcur1}-\rf{infcur2}.  
  
\subsection{A solvable model presenting toroidal solitons}
\label{sect:nicemodel}

Consider the quantity 
\be 
H_{\mu\nu} \equiv {\bf n} \cdot \( \pa_{\mu} {\bf n} \times \pa_{\nu}{\bf n}\) 
\lab{hmn}
\ee
where $\bfn$  are scalar fields living on $S^2$. Using \rf{stereo} one obtains 
\be 
H_{\mu\nu} = {-2 i \o \({1+\u2}\)^2} \, \( \pa_{\mu}u \pa_{\nu}u^* - 
\pa_{\nu}u \pa_{\mu}u^*\)
\lab{hmnu}
\ee

We introduce the Lagrangean
\be
\cl \equiv - \eta_0 \( H_{\mu\nu}^2\)^{3\o 4} = 
- \eta_0\, 8^{3\o 4} { \( K_{\mu} \pa^{\mu} u^* \)^{3\o 4} \o  
\( 1+\mid u \mid^2\)^3}
\lab{nicemodel}
\ee
where $K_{\mu}$ is the same as in \rf{kmu}, and where $\eta_0 = \pm 1$,
determines the choice of the signature of the Minkowski metric, $g_{\mu\nu}
= \eta_0 \,{\rm diag}\( 1, -1,-1,-1\)$. 

The corresponding equations of motion are
\be
\pa^{\mu} \( { \( K \pa u^* \)^{-{1\o 4}} K_{\mu} \o {1+\u2}}\) = 0
\lab{eqmot2}
\ee
and its complex conjugate.

This model possesses a representation in terms of the zero curvature \rf{lzc2}
and \rf{lzc1b}, with the potentials being given by \rf{cosetpottensor} and
\be
\ck_{\mu} \ra  { \( K \pa u^* \)^{-{1\o 4}} K_{\mu} \o {1+\u2}}
\lab{kmunice}
\ee
Indeed, such $\ck_{\mu}$ satisfies \rf{req1} and \rf{constraint} as a
consequence of \rf{reqk1} and \rf{reqk2}. 

Consequently, the model \rf{nicemodel} is integrable (or solvable) in the sense
that it possesses  an infinite number of conserved currents given by
\rf{infcur1}-\rf{infcur2}. 

We are interested in constructing exact static finite energy solutions with
non vanishing topological charges. The finite energy requirement imposes that
the field $\bfn$ should be constant at spatial infinity. Therefore, for
such purpose one can consider the three dimensional space as an $S^3$ where
the spatial infinity is identified with the north pole. The relevant
topological invariant is given by the Hopf map $S^3 \ra S^2$, and the
topological charge is 
\be
Q_h \equiv {1 \o 4 \pi^2 } \int \eps_{ijk} H_{ij} A_k d^3 x
\quad ; \quad H_{ij} = \pa_i A_j - \pa_j A_i = \eps_{abc} n^a \pa_i n^b
\pa_j n^c
\lab{hopf}
\ee

One of the main difficulties of constructing such type of solutions comes from
scaling instabilities in the energy \ct{deser,nicole,faddeev}. 
The choice of the
Lagrangean density \rf{nicemodel} is made to avoid such problems. Indeed, the
energy for static configurations is given by
\be
E \equiv \int d^3x \, \Theta_{00} = 
 8^{3\o 4}\, \int d^3x \,  { \( K_{i} \pa^{i} u^* \)^{3\o 4} \o  
\( 1+\mid u \mid^2\)^3}
\lab{energy}
\ee
with $i=1,2,3$, and $\Theta_{\mu\nu}$ being the canonical energy-momentum
tensor. Under a rescaling $x^i \ra \l x^i$, one has 
$K_{i} \ra \l^{-3}  K_{i}$, and so energy is scale invariant. 

The soliton we found is constructed using the rational map approach
\ct{HMS98}. It has a Hopf charge $Q_H=1$ and corresponds to a spherically
symmetric hedgehog Skyrme field 
defined in terms of a rational map $R :S^2 \to S^2$ and a radial 
profile function $f(r)$ which enter as follows in the expression
for the complex field $u$ 
\be
u=\h {R\o {\v R \v} } \( {\v R \v} - {1\o {\v R \v}} + 
i \( {\v R \v} + {1\o {\v R \v}} \) g(r) \) 
\lab{urat}
\ee
where
\be
g(r) \equiv {\rm cotan} f(r)
\lab{gfr}
\ee
In what follows we choose
\be
R ( \th, \phi ) = \tan (\th /2) \exp ( i \phi)
\lab{rat-map}
\ee
which can be identified via stereographic projection with a point $z$ on
the sphere defined by polar coordinates $(\th, \phi)$.
With the choice of \rf{rat-map} the complex field $u$ becomes:
\be
u = - { e^{i \phi} \o \sin (\th)} \( \cos (\th) - i g(r) \)
\quad ; \quad {1+\v u \v^2} = ( 1 + g^2 (r) ) /  \sin^2 (\th)
\lab{usub}
\ee
for which the Hopf charge $Q_H$ is equal to one \ct{BS98}.

Plugging the ansatz \rf{usub} back into equations of motion \rf{eqmot2} 
we find that it is a solution of equations of motion for 
\be
( 1 + g^2 (r) ) /r^2 = g^{\pr \, 2} (r) \quad ;\quad
g(r) = \pm (r^{-1}-r)/2
\lab{g2}
\ee
Correspondingly, the soliton solutions of equations of motion
are given by
\be
u_{\pm} = - { e^{i \phi} \o \sin (\th)} \( \cos (\th) \pm {i \o 2} 
\( {1 \o r} -r \) \)
\lab{usubpm}
\ee
According to \rf{g2}, we can take the profile function $f(r)$ to be
\be
f (r) = \arctan \( 2 r / (r^2-1) \)
\lab{g2f}
\ee
which is monotonically decreasing function with the boundary conditions
$ f (0) = \pi$ and $ f (\infty) = 0$.

We also find that the soliton energy is given by 
\br 
E= \int \( H^2_{ij} \)^{3/4} d^3 x 
= (8 \times 4^3)^{3/4} 2 \pi
\int {r^2 dr \o (r^4 +2 r^2 +1)^{3/2}}
\lab{olda}  
\er
and since $\int {r^2 dr (r^4 +2 r^2 +1)^{-3/2}} = \pi / 16$
we obtain
\be 
E = 8 (2^{3/4}) \pi^2 = 132.78
\lab{old} 
\ee
Alternatively, one can rewrite the soliton solutions \rf{usubpm} 
as a composite of the Hopf map :
\be
u_{\pm} = \pm i { \Phi_4 \pm i \Phi_3 \over \Phi_1 - i \Phi_2}
\lab{nicolu}
\ee
together with the stereographic map $:\IR^3 \to S^3$ of degree $1$:
\be
\Phi_i = { 2 x_i \over r^2 + 1} \; , i=1,2,3 \; ;\;
\Phi_4 = { r^2 -1  \over r^2 + 1}
\lab{stereoa}
\ee
Eq. \rf{usubpm} is reproduced when making use
of the spherical representation $x_1+ix_2 =r \sin \th \exp ( i \phi)$,
$x_3 = r \cos \th$. We recognize in \rf{nicolu} the soliton solution of 
reference of \ct{nicole}, where expression of the same form as in \rf{nicolu} 
was found as a solution to the equations of motion of the model:
\be
\cl = - \eta_0 \( -\eta_0 {1 \o 4} \( \pa {\bf n}\)^2 \)^{3/2}
\lab{nicol-lag}
\ee
Note, that the scaling property of this model is such that it circumvents 
the Derrick's theorem in the similar manner to the model defined by 
\rf{nicemodel}.

Equations of motion of \rf{nicol-lag} differ from that of \rf{nicemodel}.
However, when the additional constraint
\be
\( \pa u \)^2 =0
\lab{add-constraint}
\ee
is imposed then equations of motion for both models take an identical
and simplified form of
\be
\pa^{\mu} \( h\( u \) \pa_{\mu} u \) = 0
\lab{simple-eqs}
\ee
where
\be
 h\( u \) \equiv {\( \pa u \pa u^* \)^{\h } \o { 1 + \u2 }}
\ee
In the case of soliton solutions given by \rf{usub} or \rf{usubpm} the 
condition \rf{add-constraint} is automatically satisfied, which explains 
why \rf{usubpm} is a common solution for both these models.

\subsection{The Skyrme-Faddeev model}
\label{sect:sf}

The Skyrme-Faddeev (SF) model is defined by the Lagrangean \ct{faddeev} 
\be
\cl = m^2 \( \pa {\bf n}\)^2 -\eta_0 {1\o e^2} H_{\mu\nu}^2 + 
\l \( {\bf n}^2 -1\)
\lab{sflag}
\ee
where $H_{\mu\nu}$ was defined in \rf{hmn}, $\eta_0=\pm 1$ determines the
signature of the Minkowski metric (see \rf{nicemodel}), and $\l$ is a Lagrange
multiplier. 

The corresponding equations of motion, in terms of the complex field $u$
introduced in \rf{stereo}, are
\be
\( 1+\u2 \)  \pa^{\mu} L_{\mu} - 2 u^* \( L^{\mu} \pa_{\mu} u\) = 0 
\lab{eqmotsf}
\ee
and its complex conjugate, where 
\be
L_{\mu} \equiv m^2 \pa_{\mu} u - \eta_0  
{4 \o e^2}\, {K_{\mu}\o{\( 1+\mid u \mid^2\)^2}} 
\lab{lmu}
\ee
and $K_{\mu}$ is defined in \rf{kmu}.       

One of the main properties of such theory is that the two terms compensate the
scaling instabilities in the energy that each term would present if considered
separate. Indeed, the energy for static configurations is given by
\be
E \equiv \int d^3x \, \Theta_{00} = E_1 + E_2
\ee
with 
\br
E_1 &\equiv&  4 m^2 \, \int d^3 x {\mid \nabla u \mid^2 \o { \( 1 + \u2\)^2}} 
\nonu\\
E_2 &\equiv&  {32\o e^2}\, \int d^3 x { \( \nabla u_1\)^2 \( \nabla u_2\)^2 
\( 1 - \cos^2 \gamma \) \o {\( 1 + \u2\)^4}}
\er
where $\gamma$ is the angle between the vectors $\nabla u_1$ and 
$\nabla u_2$, and so  $E_1$ and $E_2$ are positive definite. 

By rescaling the space variables as $x_i \ra \l x_i$,  one has 
\be
E\( \l \) = \l E_1 + {1 \o \l} E_2
\ee
Expanding around $\l =1$ ($ \l - 1 \sim \varepsilon$)
\be
E\( \l \) =  \( E_1 + E_2 \) 
+ \( E_1 - E_2 \) \, \varepsilon 
+ E_2 \, \varepsilon^2 + O \( \varepsilon^3 \) 
\ee
one observes that it is necessary to have 
\be
E_1 = E_2 
\ee
to get stable configurations. 

However, no solution with non trivial topological charge has been 
explicitly found
for this model. The existence of soliton solutions has been 
corroborated by numerical
calculations using variational methods to find configurations which minimize
the energy \ct{faddeev,BS98}. 

We now discuss the integrability properties of the Skyrme-Faddeev model. 
Notice that 
\be
{\rm Im} \( L_{\mu} \pa^{\mu} u^* \) = 0
\ee
Therefore, if one makes the correspondence 
\be
\ck_{\mu} \ra L_{\mu}
\lab{replace}
\ee
one notices that \rf{req1} is satisfied, and the Skyrme-Faddeev  model admits
a zero curvature 
representation with the potentials being given by \rf{intpota} and
\rf{intpotb}. 

The conserved currents are given by \rf{3curr}, and they correspond to the
three Noether currents associated to the invariance of the model under the
$O(3)$ symmetry. The  Skyrme-Faddeev  model, however, does not admit an
infinite number of conserved currents, obtainable through the procedures
described in section \ref{sect:su2u1}, because the condition \rf{constraint} 
is not satisfied, since  
\be 
L_{\mu} \pa^{\mu} u = m^2 \( \pa u \)^2
\lab{ldu}
\ee

However, the submodel obtained by imposing the constraint\footnote{One could
also impose $m^2 = 0$ to obtain an integrable submodel. However, the static
solutions of such theory would present, in $3+1$ dimensions,  scaling
instabilities in the energy. To avoid such instabilities one would have to
consider such submodel in $4+1$ dimensions.}  
\rf{add-constraint}, does possess an infinite number of conserved currents 
given by \rf{infcur1}
(with the replacement \rf{replace}). Then the equations of motion \rf{eqmotsf} 
become 
\be
\pa^{\mu} \( f\( u\) \pa_{\mu} u  \) = 0
\ee 
where
\be
f\( u\) \equiv m^2 - \eta_0 
{4 \o e^2}\, {\pa_{\mu} u \pa^{\mu} u^* \o{\( 1+\mid u \mid^2\)^2}}  
\ee

Since such submodel possesses an infinite number of local conserved currents,
we believe that it is easier to be solved than the full Skyrme-Faddeev model. 
We are now investigating the  solitons it may have.  

\lskip
{\bf Acknowledgments.} H.A. gratefully acknowledges Fapesp for financial
support and IFT/UNESP for hospitality. LAF anf AHZ are partially financed by
CNPq-Brazil. 
 
\lskip

\end{document}